\documentclass[usenatbib]{mn2e}
\usepackage{graphicx}
\usepackage{ifthen}

\def\ltsima{$\; \buildrel < \over \sim \;$}
\def\simlt{\lower.5ex\hbox{\ltsima}}
\def\gtsima{$\; \buildrel > \over \sim \;$}
\def\simgt{\lower.5ex\hbox{\gtsima}}
\def\kms{{\rm\,km\,s^{-1}}}

\def\kpc{{\rm\,kpc}}

\def\AA{$\; \buildrel \circ \over {\rm A}$}

\def\UseFigs{1}

\def\deg{^\circ}

\def\s{\ifmmode \widetilde \else \~\fi}
\def\={\overline}

\def\spose#1{\hbox to 0pt{#1\hss}}

\def\lta{\mathrel{\spose{\lower 3pt\hbox{$\mathchar"218$}}
     \raise 2.0pt\hbox{$\mathchar"13C$}}}
\def\gta{\mathrel{\spose{\lower 3pt\hbox{$\mathchar"218$}}
     \raise 2.0pt\hbox{$\mathchar"13E$}}}
\def\Dt{\spose{\raise 1.5ex\hbox{\hskip3pt$\mathchar"201$}}}    
\def\dt{\spose{\raise 1.0ex\hbox{\hskip2pt$\mathchar"201$}}}    

\def\dotsfill{\leaders\hbox to 1em{\hss.\hss}\hfill}

\loadboldmathitalic 
\title[The Monoceros Ring in front of the Carina and Andromeda galaxies] {A radial velocity survey of low Galactic
latitude structures: III. The Monoceros Ring in front of the Carina and Andromeda galaxies}
\author[N. F. Martin et al.] {N. F. Martin$^{1,2}$, M. J. Irwin$^{2}$, R. A. Ibata$^{1}$, B. C. Conn$^{3}$, G. F. Lewis$^{3}$, 
\newauthor M. Bellazzini$^{4}$, S. Chapman$^{5}$ \& N. Tanvir$^{6}$\\
$^{1}$ Observatoire de Strasbourg, 11, rue de l'Universit\'e, F-67000, Strasbourg, France\\
$^{2}$ Institute of Astronomy, Madingley Road, Cambridge, CB3 0HA, U.K.\\
$^{3}$ Institute of Astronomy, School of Physics, A29, University of Sydney, NSW 2006, Australia\\
$^{4}$ INAF - Osservatorio Astronomico di Bologna, Via Ranzani 1, 40127, Bologna, Italy\\
$^{5}$ California Institute of Technology, Pasadena, CA, 91125, USA\\
$^{6}$ Physical Sciences, Univ. of Hertfordshire, Hatfield, AL10 9AB, U.K.}

\date{\today}
\begin{document} 
\maketitle 
\begin{abstract} 
As part of our radial velocity survey of low Galactic latitude structures that surround the Galactic disc, we report the
detection of the so called Monoceros Ring in the foreground of the Carina dwarf galaxy at Galactic coordinates
$(l,b)=(260\deg,-22\deg)$ based on VLT/FLAMES observations of the dwarf galaxy. At this location, 20 degrees in
longitude greater than previous detections, the Ring has a mean radial velocity of $145\pm5\kms$ and a velocity
dispersion of only $17\pm5\kms$. Based on Keck/DEIMOS observations, we also determine that the Ring has a mean radial
velocity of $-75\pm4\kms$ in the foreground of the Andromeda galaxy at $(l,b)\sim(122\deg,-22\deg)$, along with a
velocity dispersion of $26\pm3\kms$. These two kinematic detections are both highly compatible with known
characteristics of the structure and, along with previous detections provide radial velocity values of the Ring over the
$120\deg<l<260\deg$ range. This should add strong constraints on numerical models of the accretion of the dwarf
galaxy that is believed to be the progenitor of the Ring.
 
\end{abstract}
  

\section{Introduction}

With the public release of all sky surveys, many details have been gained in the structure of the outer parts of the
Galactic disc. In particular, the Sloan Digital Sky Survey (SDSS) revealed the existence of a stellar structure
in the anticentre direction, near the Galactic plane and slightly over the edge of the disc \citep{newberg02}. Visible
as a clear main sequence at a Galactocentric distance of $18\kpc$ for $180\deg<l<225\deg$ and $|b|<30\deg$, this
structure is unlike what is expected for the Galactic disc. \citet{ibata03} used the INT Wide Field Camera to show
this structure is in fact present in the second and third Galactic quadrants, circling the disc in a ring-like fashion.
Radial velocity measurements also revealed a kinematically cold population with a velocity dispersion of only
$15-25\kms$, once more unexpected for a Galactic structure, leading to the conclusion that this so-called
Ring\footnote{This structure has been called many names including the Monoceros Ring, the Galactic Anticentre Stellar
Structure or GASS and the Ring. Given its extent in Galactic longitude and its presence in numerous constellations, we
prefer to call it simply the Ring.} is produced by the accretion of a dwarf galaxy in the Galactic plane whose
tidal arms are wrapped around the Milky Way.

Subsequently, much work has been invested in trying to determine the true extent of this Ring and where possible
determine its kinematics to constrain the orbit of its progenitor. \citet{rocha-pinto03} used the 2MASS catalogue to probe the
distribution of M giant stars and found that the Ring may extend over the $120\deg\simlt l\simlt270\deg$ range in the
Northern hemisphere and be present in the Southern hemisphere, with $|b|<35\deg$. \citet{conn05a}
extended the \citet{ibata03} INT/WFC survey to show the Ring is present within 30 degrees of the Galactic plane
throughout the whole second quadrant but seems to disappear at $l\sim90\deg$. More constraints on the structure
were gained by the \citet{crane03} spectroscopic survey of putative Ring M-giant stars in the anticentre direction
($150\deg<l<230\deg$, $|b|<40\deg$) which confirmed that most of these stars are indeed linked to this structure. Using
the simple model of a population orbiting the Milky Way in a prograde, circular orbit, they determined their sample was
best fitted by a population at a Galactocentric distance of $18\kpc$ and with a rotational velocity of $220\kms$. The
resulting velocity dispersion of $20\pm4\kms$ around this model is in good agreement with the velocity dispersion
measured by the SDSS team \citep{yanny03}.

While tracking the Ring, two other structures were discovered that do not seem to be directly related to the Monoceros
Ring. Using the 2MASS catalogue, \citet{rocha-pinto04} reported the existence of a diffuse population of M giants in the
direction of the Triangulum and Andromeda constellations, behind known detections of the Ring. Also using the 2MASS
catalogue, our group presented evidence of a dwarf galaxy located closer than the Ring, just at the edge of the Galactic
disc, in the Canis Major constellation \citep{martin04,bellazzini04}. In particular, we argued that the accretion of the
Canis Major galaxy onto the Galactic disc would naturally reproduce similar features as the one observed for the Ring.
However, a link between these three structures remains putative at the moment, even though current models show such
a scenario is highly plausible \citep{penarrubia05,martin05,dinescu05}.

To gain more insight into the nature of these structures, we have started a radial velocity survey of regions at low
Galactic latitude where the Ring structure may be located. Using the AAT/2dF multi-fibre spectrograph, we first
targeted the Canis Major dwarf \citep{martin05}. This survey also revealed the presence of the Ring behind the dwarf,
with a Heliocentric radial velocity of $133\pm1\kms$ and dispersion of $23\pm2\kms$ at a Galactocentric distance of
$19\kpc$, highly compatible with previous detections \citep{conn05b}. In this third paper of the series, we report
the presence of the Ring in the foreground of the Carina dwarf galaxy from VLT/FLAMES observations of the dwarf at
$(l,b)=(260\deg,-22\deg)$. Using Keck2/DEIMOS observations of regions around M31, we also determine the radial velocity
of the Ring at $(l,b)\sim(122\deg,-22\deg)$ where the Ring is known to exist but where its radial velocity has not yet
been measured. Section 2 presents the Ring detection in front of Carina and Section 3 deals with the detection in
front of the Andromeda galaxy. Section 4 concludes this letter.

In the following, all the magnitudes have been corrected for extinction using the maps from
\citet{schlegel98}. We also assume that the Solar radius is $R_\odot = 8\kpc$, that the LSR circular velocity is
$220\kms$, and that the peculiar motion of the Sun is ($U_0=10.00\kms, V_0=5.25\kms, W_0=7.17\kms$; \citealt{dehnen98}).
Except when stated otherwise, the radial velocities, $v_{r}$, are Heliocentric radial velocities, not corrected for the
motion of the Sun.

\section{The Ring in front of the Carina dwarf galaxy}

FLAMES observations in the Carina fields were taken from the public access ESO raw data
archive\footnote{http://www.eso.org/archive} and included 14 setups centred on various fields designed to study the
Carina dwarf galaxy. The total integration time of each setup was typically $15000$ seconds giving excellent
signal-to-noise ($>$20:1 per 0.2\AA\ sampling element) for the Galactic foreground stars in these directions.  The
Carina data plus calibration sequences were downloaded and processed through the standard ESO FLAMES low resolution
pipeline.  

At the time of using the pipeline, sky subtraction was not part of the pipeline procedure so the remaining processing
steps used the FLAMES software developed for the DART project (see e.g. \citealt{tolstoy04}). Briefly, this
software stacks the individual repeat spectra on the same target field; combines all the sky observations to form a
master sky spectrum; and then optimally scales, shifts (if necessary) and resolution-matches the sky to the object
spectrum prior to sky subtraction.  The sky-subtracted spectra are then searched for CaII near infrared triplet lines
and a model template CaII spectrum is used to extract velocity information.

The direct imaging used here also came from the ESO archive and comprised 4 $V,I$ sequences of ESO/WFI data covering
$\approx 1 \times 1$ square degrees centred on the Carina dwarf.  This was processed through the standard Cambridge
pipeline \citep{irwin01} to produce catalogues of magnitudes, colours and object morphological classifications.

\begin{figure}
\ifthenelse{\UseFigs=1}{
\includegraphics[angle=270, width=\hsize]{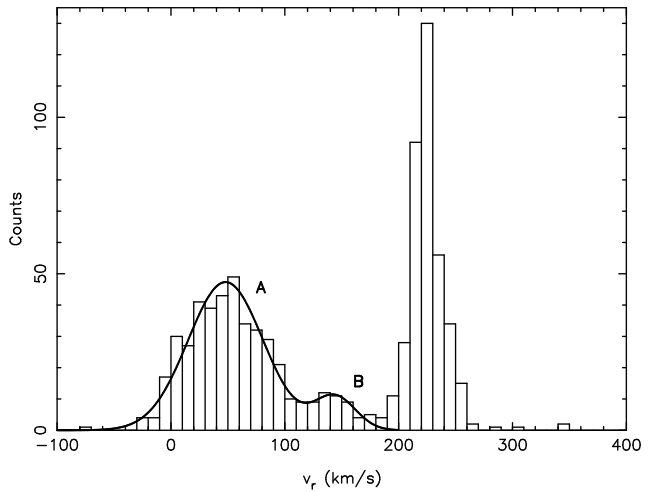} }{caption1}
\caption{Velocity distribution of the FLAMES sample. The peak at $v_{r}\sim220\kms$ is produced by Carina stars. Stars
with $v_{r}<180\kms$ are well fitted by a double Gaussian model (represented by the black thick line) with populations
A and B centred on $49\pm7$ and $145\pm5\kms$ and with dispersions of $33\pm2$ and $17\pm5\kms$ respectively.}
\end{figure}

The velocity distribution of all the stars targeted with FLAMES is displayed on Figure~1. The most prominent feature is
of course the peak of stars at $v_{r}\sim220\kms$ produced by Red Giant Branch stars belonging to the Carina dwarf
galaxy (the primary targets of these FLAMES observations). Stars with $v_{r}<180\kms$ are expected to be Galactic stars
belonging to the thin disc, thick disc and/or stellar halo. However, they seem to follow a bimodal Gaussian distribution
produced by populations with central velocities and dispersions of $(\mu_A,\sigma_A)=(49\pm7\kms,33\pm2\kms)$ and
$(\mu_B,\sigma_B)=(145\pm5\kms,17\pm5\kms)$ and a 9 to 1 ratio according to a  maximum likelihood fit of a two component
Gaussian model. Although population A has the expected characteristics of a disc-like population in the foreground, the
low dispersion of population B is more puzzling.

\begin{figure}
\begin{center}
\ifthenelse{\UseFigs=1}{
\includegraphics[angle=0,width=\hsize]{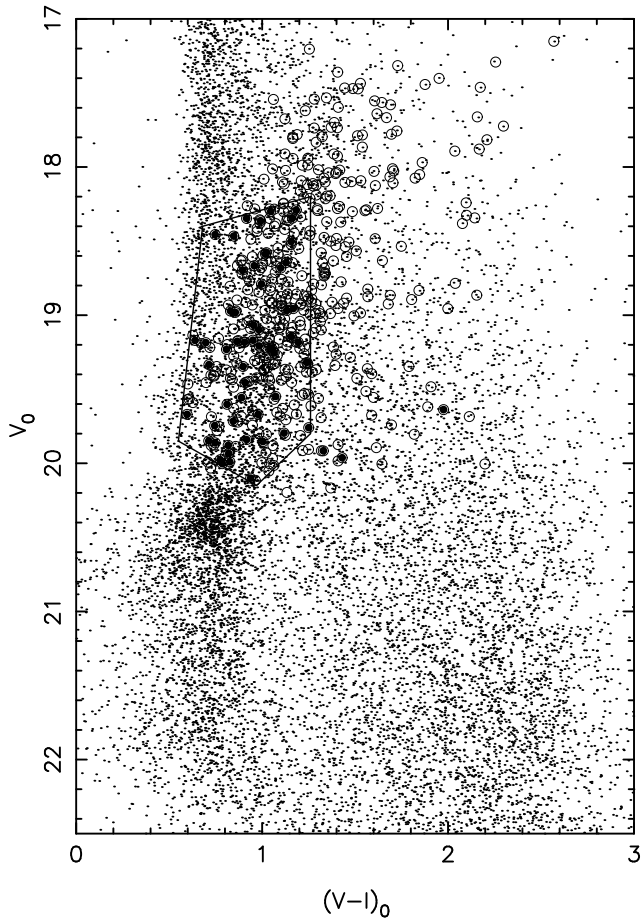} }{caption1}
\caption{ESO/WFI Colour-Magnitude Diagram of the Carina region. Red Clump stars from the dwarf produce the
feature around $V_{0}\sim20.5$ and $(V-I)_{0}\sim0.7$. DEIMOS target stars, that where chosen along the dwarf's red giant
branch, are highlighted with those that show a radial velocity under $110\kms$ plotted as hollow circles (population
A) while stars with $110<v_{r}<180\kms$ are shown as filled circles (population B). The two populations
show highly dissimilar behaviours, with almost all population B stars bluer than $(V-I)_0=1.3$. The selection box that is
used for comparison with the Besan\c{c}on model is also shown.}
\end{center}
\end{figure}

The Colour-Magnitude Diagram (CMD) of Figure~2 displays the location of stars belonging to these two populations with
hollow circles for population A stars ($v_{r}<110\kms$) and filled circles for population B stars ($110<v_{r}<180\kms$).
The two populations show drastically different colour-magnitude distributions, pointing at a genuine difference between
them. Population A is widely spread in colour and follows, once more, the expected distribution of foreground disc
stars. On the other hand population B is confined on the bluer part of our sample, with almost all stars having
$(V-I)_0<1.3$. Of the three stars with $(V-I)_0>1.3$, the one with the highest $(V-I)_0$ is only just a population B
star with $v_{r}=112\kms$ (the two others have radial velocities of 143 and $149\kms$).

Such a sharp colour cut makes it very unlikely that population B is produced by disc stars. Among the Galactic
components, only the stellar halo is expected to have such behaviour (see e.g. \citealt{conn05a} for a comparison of the
disc and halo Galactic component CMDs as they appear in the Besan\c{c}on model of \citealt{robin03}). However, the
velocity dispersion of the halo is $\sim100\kms$ \citep[e.g.][]{gould03}, at odds with the $17\kms$ found for population
B. Since this low dispersion could be an artifact of the radial velocity cuts we used, we investigate the number of
stars that would be expected for a halo population at $(l,b)=(260\deg,-22\deg)$. According to the synthetic Besan\c{c}on
model of Galactic stellar populations \citep{robin03}, there should be less than 225 halo stars per square degree in the
CMD region highlighted on Figure~2 and that contains most of our population B stars, independently of radial velocity.
Since our FLAMES targets roughly cover 0.25 square degrees, we would expect our sample to contain less than 60 halo
stars if complete. Assuming the Milky Way is surrounded by a non-rotating stellar halo with a velocity dispersion of
$100\kms$, our sample should contain only $\sim15$ halo stars within $110\kms<v_{r}<180\kms$, once again, if it were
complete. Since within the selection box of Figure~2, the completeness is under 5\%, it is highly unlikely that the 56
population B stars belong to the halo. Therefore, population B is most likely a non-Galactic population of stars that
lie in front of the Carina dwarf galaxy, with a mean radial velocity of $145\pm5\kms$ and a velocity dispersion of
$17\pm5\kms$.

A direct comparison with the velocity distribution of all the stars  from the model that fall in the same region of the
CMD would of course be more suited for our purpose. However the model greatly overpredicts the number of thick disc
stars in this region of the sky when compared to the observations. This is probably due to an overestimate of the thick
disc flare in the model. Although population A is very well reproduced by the thin disc population of the model, the
modeled thick disc population is twice as numerous, centred on $v_{r}\sim80\kms$ and with a high dispersion of
$\sim50\kms$. Such a population is clearly not present in our data. Since the sharp $(V-I)_0<1.3$ colour limit of
population B is not expected for a disc population, we prefer comparing population B with only the halo population of
the model.

With a detection of the Ring behind the Canis Major galaxy under the Galactic disc only 20 degrees away in Galactic 
longitude from the feature we detect in front of Carina, this population could be another detection of the Ring. 
Previous detections of the Ring have mainly relied on CMDs and especially on the main sequence of
this population compared to the disc population at brighter magnitudes \citep{newberg02,ibata03,conn05a}. Unfortunately,
for the Carina CMD, the red clump of the dwarf galaxy at $(V-I)_0\sim0.7$ and $V_0\sim20.5$ lies in the region of
interest for detecting the Ring main sequence. However, comparison with fiducials shows the location of population B
stars is not incompatible with turn-off stars from the old metal-rich population at a Galactocentric distance of
$\sim20\kpc$ that is usually assumed for the Ring; even though the Carina-driven selection criteria applied to select
the stars in the sample prevents a reliable comparison.

The radial velocity characteristics of this population in front of Carina further supports a connection with the Ring.
Indeed, the determined velocity dispersion of $17\pm5\kms$ is within the $15-25\kms$ range of the SDSS and
\citet{crane03} detections and close to the $24\pm2\kms$ of the detection in the background of Canis Major. For this
latter case, the higher uncertainties on the 2dF radial velocity value of each star ($\sim10\kms$) compared to those of
the FLAMES derived velocities ($\sim3\kms$) may also artificially increase the dispersion. When corrected from the Solar
motion, the mean radial velocity of population B, $v_{\mathrm{gsr,B}}=-65\kms$, is very close to the
$v_{\mathrm{gsr,bCMa}}=-67\kms$ found only 20 degrees away behind the Canis Major dwarf galaxy.

Therefore, we conclude that the non-Galactic population we have uncovered in front of the Carina dwarf galaxy is most
likely part of the Ring.

\section{Kinematics of the Ring in front of the Andromeda galaxy}

\begin{figure}
\ifthenelse{\UseFigs=1}{
\includegraphics[angle=270,width=\hsize]{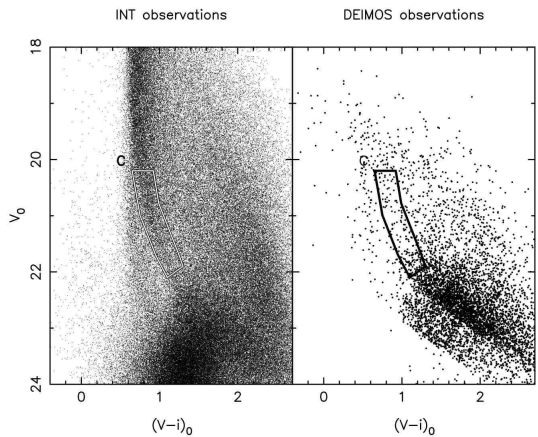} }{caption1}
\caption{Colour-Magnitude Diagram of the INT `M31-N' field of \citet{ibata03} (left panel) and the corresponding stars
targeted with DEIMOS (right panel). The clear main sequence of the INT CMD is used to define a selection box of Ring
stars (labeled C).}
\end{figure}

\begin{figure}
\ifthenelse{\UseFigs=1}{
\includegraphics[angle=270,width=\hsize]{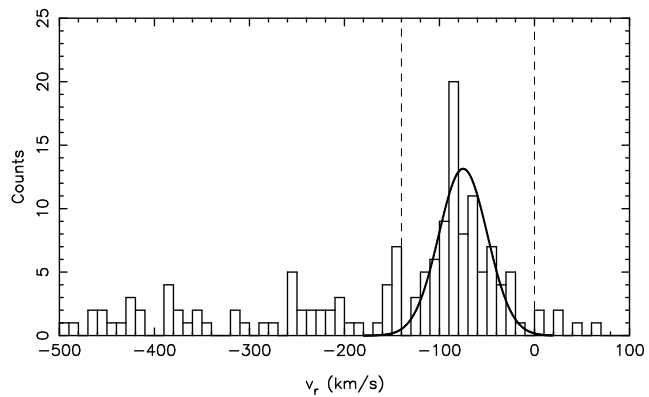} }{caption1}
\caption{Radial velocity distribution of DEIMOS stars that fall in the Ring selection box (labeled C in Figure~3). The
few stars with $v_{r}\sim-200\kms$ may contain M31 disc stars (the primary targets of these observations). Observed stars
with $-140\kms<v_{r}<0\kms$ (within the two dashed lines) are well fitted by a Gaussian model with a mean velocity of
$-75\pm4\kms$ and a dispersion of $26\pm3\kms$ (represented by the black thick line).}
\end{figure}

The presence of the Ring in front of the Andromeda galaxy was first reported by \citet{ibata03} from the analysis of
INT/WFC Colour-Magnitude diagrams. The CMD of their `M31-N' field is shown on the left panel of Figure~3 and the Ring is
clearly visible as a main sequence that extends from $(V-i,V)_{0}\sim(1.2,22)$ to $(V-i,V)_{0}\sim(0.6,20.0)$. During
our Keck/DEIMOS survey of M31 outer disc and halo substructures \citep[e.g.][]{ibata05}, we took the opportunity to
target foreground Ring stars that fortuitously fall in the targeted regions. The data were reduced as in \citet{ibata05}
and the CMD of all the stars with a radial velocity uncertainty lower than $10\kms$ is presented on the right panel of
Figure~3. To select only probable members of the Ring structure, we construct a selection box around the Ring main
sequence from the INT CMD (box C in Figure~3). Among the stars in the DEIMOS sample, 86 fall within this Ring selection
box and the radial velocity distribution of these stars is displayed on Figure~4. Aside from Galactic halo and M31 disc
stars at $v_{r}<-200\kms$, a peak is apparent at $\sim-70\kms$. Applying a maximum likelihood algorithm to fit with a
Gaussian model those stars with $-140<v_{r}<0\kms$ that produce the peak, reveals that this population has a mean
velocity of $-75\pm4\kms$ and an intrinsic dispersion of $26\pm3\kms$, corrected for the uncertainties on each measured
radial velocity.

A direct comparison with the radial velocity distribution of stars from the Besan\c{c}on model within the same sky
region ($119\deg<l<124\deg$ and $-24\deg<b<-19\deg$) and that fall in the same CMD selection box reveals the Ring
sub-sample is not incompatible with the model. Indeed, a Kilmogorov-Smirnov test yields a probability of 10 percent that
the two populations are identical. However, it is not unexpected that disc stars and Ring stars should show similar
behaviour since both populations are believed to orbit the Milky Way on nearly circular orbits and the small shift in
distance between them does not translate into a significant difference in radial velocity. Given that our
selection box is constructed to contain Ring stars, it would be surprising that all the stars of the sample belong to the
Galactic disc. On the contrary, we find it more likely that we are observing mainly Ring stars, as is suggested by the
relatively low velocity dispersion of $26\pm3\kms$ in the sample, at odds with the $\sim50\kms$ found in the
Besan\c{c}on model within the selection box. This low dispersion is also compatible with previous detections, especially
since some disc and/or halo stars certainly fall in the same radial velocity range and increase the dispersion. The mean
Heliocentric velocity of $-75\pm4\kms$ which converts to a Galactocentric standard of rest radial velocity of
$v_{\mathrm{gsr}}=94\pm4\kms$ is also only slightly higher than the simple circular \citet{crane03} model. As for the
detection in the foreground of the Carina dwarf, the radial velocity similarities between the previous detections of the
Ring and the population in front of M31 strengthen the Ring nature of our detection.

\section{Summary and Conclusion}
\begin{figure}
\ifthenelse{\UseFigs=1}{
\includegraphics[width=\hsize]{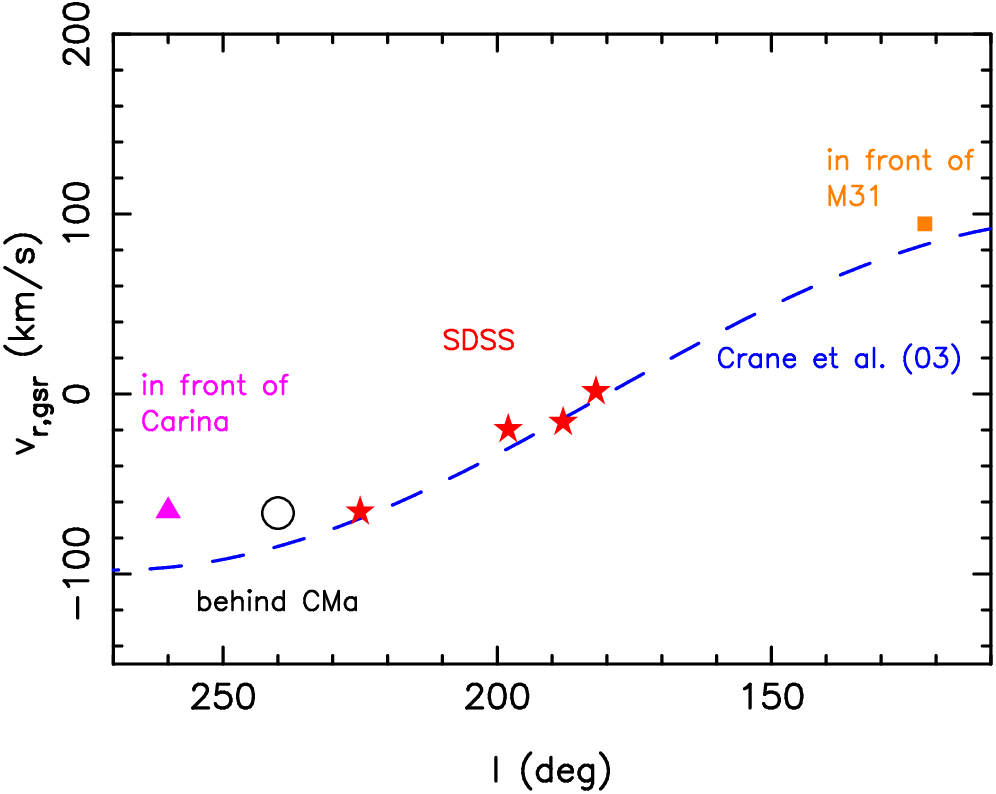} }{caption1}
\caption{Summary of all known radial velocity detections of the Ring in the Galactic standard of rest. The dashed line
corresponds to the \citet{crane03} model of the Ring: a population orbiting the Milky Way at a Galactocentric distance
of $18\kpc$ and with a rotational velocity of $220\kms$. The \citet{yanny03} SDSS detections are shown as stars, the
\citet{conn05b} detection behind Canis Major is shown by a hollow circle and the two detections of this letter in front
of Carina and M31 are shown as a filled triangle and a filled square respectively. Uncertainties on these values are
not shown since they are smaller than the used symbols. Though near the \citet{crane03} model, detections away from the
anticentre tend to be offset from the model.}
\end{figure}

We have presented the detection of two groups of stars that lie in front of the Carina dwarf galaxy at
$(l,b)=(260\deg,-22\deg)$ and in front of the Andromeda galaxy at $(l,b)\sim(122\deg,-22\deg)$ and that cannot be
satisfyingly explained by known Galactic components. The proximity with known detections of the Ring that surrounds the
Galactic disc makes it highly probable that they belong to the same structure. Both detections have a low velocity
dispersion ($17\pm5\kms$ and $26\pm3\kms$ respectively), a characteristic value encountered in all previous detections of
the Ring.

With the Ring detection behind the Canis Major dwarf galaxy reported by \citet{conn05b}, the radial velocity of the Ring
population is now sampled throughout the $120\deg<l<260\deg$ range, which should provide important constraints on
N-body models. However, it can be directly seen in Figure~5 that currently known radial velocity values for the Ring are
not exactly reproducible by the \citet{crane03} simple circular model. In fact, trying to fit all the detections in a
single orbit of the progenitor proves unsatisfactory, whether the orbit is forced to be circular or allowed to be
slightly elliptical. It would therefore seem that models where the Ring completely surrounds the Galactic disc with
multiple tidal arms are following the right track (see e.g. \citealt{penarrubia05} and \citealt{martin05}). In
addition to the new radial velocities we report in this letter, such models would highly benefit from a similar survey
as the one presented in \citet{conn05a}, but this time to higher longitudes to study in more detail the morphology of
the Ring in these regions and especially to add a distance constraint to the detection in front of the Carina dwarf.

\section*{acknowledgements}
NFM is grateful to the IoA for the kind hospitality during the months at Cambridge in which this work was mainly performed.
NFM acknowledges support from a Marie Curie Stage Research Training Fellowship under contract MEST-CT-2004-504604.
GFL acknowledges support from ARC DP 0343508 and is grateful to the Australian Academy of Science for financially
supporting a collaboratory visit to Strasbourg Observatory.

\newcommand{\mnras}{MNRAS}
\newcommand{\pasa}{PASA}
\newcommand{\nat}{Nature}
\newcommand{\araa}{ARAA}
\newcommand{\aj}{AJ}
\newcommand{\apj}{ApJ}
\newcommand{\apjl}{ApJ}
\newcommand{\apjs}{ApJSupp}
\newcommand{\aap}{A\&A}
\newcommand{\aaps}{A\&ASupp}
\newcommand{\pasp}{PASP}

\end{document}